\documentclass[showpacs,amsmath,amssymb, prb,twocolumn]{revtex4}
\usepackage{graphicx}
\usepackage{dcolumn}
\usepackage{float}
\usepackage{bm}
\begin{document}

\title{On the Bardeen-Cooper-Schrieffer interaction in quantum graphs}

\author{Francesco Romeo$^{1,2, \ast}$}
\affiliation{$^{1}$Dipartimento di Fisica ''E. R. Caianiello'', Universit\`a degli Studi di Salerno,
Via Giovanni Paolo II, I-84084 Fisciano (Sa), Italy\\
$^{2}$INFN, Sezione di Napoli, Gruppo collegato di Salerno,
Via Giovanni Paolo II, I-84084 Fisciano (Sa), Italy\\
$^{\ast}$email: fromeo@sa.infn.it (corresponding author)\\
\texttt{http://orcid.org/0000-0001-6322-7374}}

\begin{abstract}
We introduce a real-space version of the Bardeen-Cooper-Schrieffer interaction allowing the investigation of the non-trivial interplay between many-body physics and particles confinement on a quantum graph. When the two-body problem is considered, we find that the two-particle wavefunction is solution of an integro-differential Schr\"{o}dinger equation. The solution of the two-body eigenproblem shows the presence of a two-particle bound state whose stability is enhanced in graphs with peculiar topology. We demonstrate that the enhancement effect is robust against many-body effects, which can be studied by means of the Richardson exact solution of the many-body problem. These findings suggest that the effective pairing interaction can be enhanced in quantum graphs with appropriate connectivity. Experimental evidences in Josephson junctions arrays are also discussed in connection with the microscopic mechanism described in the present work.
\end{abstract}
\maketitle
\section{Introduction}
The interaction between fermionic particles is a central research topic in theoretical physics and it is encountered in different contexts ranging from the quark structure of baryons\cite{nuclearbook} to the superfluid states of the neutron stars\cite{neutronstars}. In condensed matter physics, Mott insulators\cite{mott} and superconductors\cite{degennesbook} are prototype systems in which many-body interactions induce completely new orders.\\
The superconducting order \cite{onnes}, in particular, elucidated by theoretical contributions due to Bardeen, Cooper and Schrieffer \cite{cooper,bcs}(BCS), originates from the formation of paired fermions stabilized by a mediated attractive interaction. The formation mechanism of two-particle bound states in a bulk metal has been clearly shown by Cooper \cite{cooper} and, after him, these composite objects are known as Cooper pairs. After the Cooper seminal work, several variants of the Cooper problem have appeared in literature \cite{gulacsi,dellano,lages,martikainen,croitoru} also aiming at exploring the interplay between interaction, confinement and structured environments \cite{croitoru,martikainen}.\\
While the problem of the non-trivial interplay between interaction and dimensionality is well-known in physics, less is known about the physics of interacting quantum systems on graph-like networks.\\
The latter ingredient has been considered in a pioneering work [\onlinecite{burioni1}] by Burioni and coworkers. These authors have demonstrated that spatial Bose-Einstein condensation can occur in dimension $d < 2$ when special discrete lattices are considered. The condensation of free bosons in these discrete structures is originated by an effective interaction induced by the network topology. In particular, it has been demonstrated that network nodes with higher connectivity act as localization centres for the bosons density [\onlinecite{burioni2,burioni3}].\\
Within the context of superconductivity, the interplay between interaction and connectivity has been clearly addressed by DeGennes and Alexander [\onlinecite{micronetworkbook}] who formulated a micronetwork theory aiming at describing situations in which the superconducting state nucleates inside an insulating matrix. The resulting theory, consisting in a Ginzburg-Landau approach on graphs, shows that the superconducting transition temperature is affected by the network connectivity. A discrete version of the DeGennes-Alexander theory has been recently formulated to describe the connection of superconducting islands coupled by Josephson tunnel junctions [\onlinecite{micronetdiscreto}]. In the latter case, the discreteness of the structure stabilizes a localized solution of the order parameter accompanied by a critical temperature enhancement. This conclusion is compatible with a recent theoretical work\cite{gastiasoro} showing that, under appropriate circumstances, non-magnetic disorder can be beneficial for the stabilization of the superconducting phase. In this respect, superconductivity in discrete graphs appears to be a model able to capture the percolative nature of the superconducting state in systems affected by disorder. From the conceptual viewpoint, these systems are formed by superconducting puddles arranged into the shape of a quantum graph\cite{notaQG}, so that the system connectivity is much more relevant than the real dimensionality.\\
Inspired by results in Ref. [\onlinecite{burioni1}], in the last decades superconducting networks shaped in the form of star or double comb have been realized and characterized\cite{silvestrini,lorenzo,ottaviani,lucci1,lucci2,lucci3}. The experimental evidence shows that the network connectivity plays a crucial role for the phase transition so that superconductivity can be enhanced in these networks\cite{lucci3}.\\
Since connectivity is an important driver for the phase transition of graph-shaped systems\cite{micronetdiscreto,deluca}, new physics is expected to arise in these synthetic structures. In order to capture these exotic effects and gain a deep understanding of the problem, one needs to formulate a genuine many-body theory able to evidence emerging features which could be missed within a mean field approach. To proceed further with this program, the formulation of the BCS problem in real space is required, being the latter one of the main achievements of the present work. By using a variety of methods, we will demonstrate that the effective interaction strength between fermions is enhanced in quantum graphs with appropriate connectivity. These findings appear to be relevant either from the fundamental viewpoint and for the emerging field of quantum technologies.\\
The work is organized as follows. In Sec. \ref{sec:pairing}, we propose a real-space version of the BCS pairing interaction which allows to study the interplay between pairing and network topology. We demonstrate that the resulting many-body Hamiltonian can be diagonalized by using the Richardson procedure. In Sec. \ref{sec:twob}, we study the two-particle problem also demonstrating that the wavefunction is solution of an integro-differential equation. Solutions of the mentioned equation are used to evaluate the two-particle bound state stability, which is affected by the network topology. The many-body problem is studied in Sec. \ref{sec:many}. Conclusions are reported in Sec. \ref{sec:concl}.

\section{Pairing interaction in quantum graphs}
\label{sec:pairing}
We are interested in describing BCS-like pairing in quantum systems lacking of translational invariance. Such structures allow to study the interplay between peculiar confinement effects and many-body physics and mimic the percolative structure of the superconducting order nucleating inside an insulating matrix. Graph and tree-like networks (i.e., connected, undirected, acyclic graphs) provide a schematic model of these systems. Thus, hereafter, we introduce a BCS-like pairing interaction in real space.

\subsection{BCS pairing interaction in real space}
In order to provide a gentle introduction to this topic, let us start with the Hamiltonian of a translational invariant system in one dimension. In momentum space, interaction between fermions is described by the Hamiltonian:
\begin{eqnarray}
\label{eq:Hk}
H=\sum_{k\sigma}\epsilon_k c^{\dagger}_{k\sigma}c_{k \sigma}-g \sum_{kk'}c^{\dagger}_{k\uparrow}c^{\dagger}_{-k \downarrow}c_{-k'\downarrow}c_{k'\uparrow},
\end{eqnarray}
where $\sigma \in \{\uparrow,\downarrow\}$ represents the particles spin projection, $k, k'$ are momentum quantum numbers, $g>0$ represents the BCS pairing strength, while creation/annihilation $c^{\dagger}_{k\sigma}/c_{k \sigma}$ operators obey standard anticommutation relations. In order to write the model in real space, let us assume that the single particle term of the Hamiltonian originates from the real-space hopping Hamiltonian:
\begin{eqnarray}
\label{eq:H0}
H_0=-K \sum_{j\sigma}c^{\dagger}_{j+1\sigma}c_{j\sigma}+h.c.
\end{eqnarray}
describing the particles hopping on an $N$-sites one-dimensional lattice with periodic boundary conditions. Accordingly, fermionic operators in real space can be expressed in terms of the single-particle wavefunctions as follows:
\begin{eqnarray}
\label{eq:cj}
c_{j\sigma}=\frac{1}{\sqrt{N}}\sum_{k}c_{k\sigma}e^{ikj}.
\end{eqnarray}
Thus, using Eq. (\ref{eq:cj}) in (\ref{eq:H0}), we get the single-particle term contributing to the Hamiltonian written in Eq. (\ref{eq:Hk}). The relation in Eq. (\ref{eq:cj}) can be inverted so that
\begin{eqnarray}
\label{eq:ck}
c_{k\sigma}=\frac{1}{\sqrt{N}}\sum_{\ell}c_{\ell \sigma}e^{-ik \ell},
\end{eqnarray}
being $\ell$ a site index and $k$ the momentum quantum number. Using Eq. (\ref{eq:ck}) in (\ref{eq:Hk}), we get:
\begin{eqnarray}
\label{eq:Hrs}
H=H_0-g\sum_{\ell r}c^{\dagger}_{\ell\uparrow}c^{\dagger}_{\ell \downarrow}c_{r\downarrow}c_{r\uparrow},
\end{eqnarray}
with $H_0$ given in Eq. (\ref{eq:H0}) and $\ell, r$ site indices. Interestingly, BCS pairing appears to be non-local in real space and rather different from the (attractive) Hubbard interaction. Despite the apparent difference, BCS and attractive Hubbard interaction can be written according to the general expression:
\begin{eqnarray}
\label{eq:HI}
H_I=-g\sum_{\ell r}\Gamma_{\ell r} c^{\dagger}_{\ell\uparrow}c^{\dagger}_{\ell \downarrow}c_{r\downarrow}c_{r\uparrow},
\end{eqnarray}
with $\Gamma_{\ell r}=1$ for BCS interaction or $\Gamma_{\ell r}=\delta_{\ell r}$ for the attractive Hubbard model. Interestingly, the pairing term in Eq. (\ref{eq:Hrs}) appears to be a completely nonlocal version of the \textit{pair-hopping term} considered in Eq. (1) of Ref. [\onlinecite{dolcini1,dolcini2}].

\subsection{Pairing Hamiltonian for quantum graphs}
Hereafter, we develop a model of fermions interacting via BCS-like interaction and constrained to move in quantum graphs. Despite these systems lack of translational invariance, it is reasonable to assume that BCS pairing preserves the same form given in Eq. (\ref{eq:Hrs}). In this way, a general pairing model is described by the Hamiltonian:
\begin{eqnarray}
\label{eq:Htree}
\mathcal{H}=\sum_{ij\sigma}c^{\dagger}_{i\sigma}h_{ij}c_{j\sigma}-g\sum_{\ell r}\Gamma_{\ell r} c^{\dagger}_{\ell\uparrow}c^{\dagger}_{\ell \downarrow}c_{r\downarrow}c_{r\uparrow},
\end{eqnarray}
with $h_{ij}$ an Hermitian matrix specifying the onsite potentials and the hopping terms connecting adjacent sites. In order to study the interplay between connectivity and many-body interaction, we confine our attention to the case in which $h_{ij}=\epsilon_{i}\delta_{ij}-K \mathcal{A}_{ij}$, where $\epsilon_{i}$ is the onsite potential, $K>0$ represents the hopping integral, while $\mathcal{A}_{ij}$ is the adjacency matrix. Adjacency matrix is a real and symmetric matrix presenting vanishing diagonal elements (i.e., $\mathcal{A}_{ii}=0$). Moreover, $\mathcal{A}_{ij}=1$ when hopping is allowed between the lattice sites $i$ and $j$, while $\mathcal{A}_{ij}=0$ for disconnected sites.\\
BCS pairing of particles constrained to move in graph-like structures is thus described by Eq. (\ref{eq:Htree}) with $\Gamma_{\ell r}=1$, i.e.
\begin{eqnarray}
\label{eq:HtreeBCS}
\mathcal{H}=\sum_{ij\sigma}c^{\dagger}_{i\sigma}h_{ij}c_{j\sigma}-g\sum_{\ell r} c^{\dagger}_{\ell\uparrow}c^{\dagger}_{\ell \downarrow}c_{r\downarrow}c_{r\uparrow}.
\end{eqnarray}
Hereafter, we demonstrate that the latter many-body problem admits an exact solution in terms of Richardson's ansatz \cite{richardson1,richardson2}. To prove the latter statement, Eq. (\ref{eq:HtreeBCS}) is rewritten in terms of fermionic fields $a_{i\sigma}$ diagonalizing the non-interacting part of the Hamiltonian. Using the unitary transformation $c_{i\sigma}=\sum_{j}U_{ij}a_{j\sigma}$, one obtains:
\begin{eqnarray}
\sum_{ij\sigma}c^{\dagger}_{i\sigma}h_{ij}c_{j\sigma}=\sum_{i\sigma}E_i a_{i\sigma}^{\dag}a_{i\sigma},
\end{eqnarray}
with $E_i$ the i-th single-particle energy level. Once the pairing part of the Hamiltonian has been expressed in terms of new fermionic fields, the complete Hamiltonian takes the following form:
\begin{eqnarray}
\mathcal{H}=\sum_{i\sigma}E_i a_{i\sigma}^{\dag}a_{i\sigma}-g \sum_{ijkl}V_{ij}^{\ast}V_{kl}a^{\dagger}_{i\uparrow}a^{\dagger}_{j \downarrow}a_{k\downarrow}a_{l\uparrow},
\end{eqnarray}
where $i,j,k,l$ are indices labeling the single particle energy levels, while $V_{ij}=\sum_{k}U_{ki}U_{kj}$. Interestingly, $h_{ij}$ matrix, being real and symmetric, admits real eigenvectors so that $U_{ij}=U^{\ast}_{ij}$. Thus, one obtains $\delta_{ij}=\sum_{k}U^{\ast}_{ki}U_{kj}=\sum_{k}U_{ki}U_{kj}=V_{ij}$, being the latter consequence of the unitary condition of the fields' transformation. In this way, we obtain the model:
\begin{eqnarray}
\label{eq:reducedBCS}
\mathcal{H}=\sum_{i\sigma}E_i a_{i\sigma}^{\dag}a_{i\sigma}-g \sum_{ij}a^{\dagger}_{i\uparrow}a^{\dagger}_{i \downarrow}a_{j\downarrow}a_{j\uparrow},
\end{eqnarray}
showing the tendency to create pairs of particles with opposite spin projection and equal single-particle energy $E_i$. Moreover, the connectivity of the graph-like network is encoded inside the single-particle energy spectrum $E_i$. A comparison between Eq. (\ref{eq:HtreeBCS}) and (\ref{eq:reducedBCS}) shows that the interaction part of the Hamiltonian maintains its structure under unitary transformation.\\
Surprisingly, the Hamiltonian model in Eq. (\ref{eq:reducedBCS}), already known in literature as \textit{reduced BCS model}, is commonly used to describe ultrasmall superconducting grains \cite{sierra,mastellone} and the related many-body problem admits exact solution as shown by Richardson.\\
The aforementioned arguments suggest that Eq. (\ref{eq:HtreeBCS}) provides an appropriate description of the BCS pairing in real space.

\section{Two-body problem}
\label{sec:twob}
The two-body problem is conveniently addressed adopting a first quantization formalism\cite{weisz,claro,souza,longhi}. In order to proceed along this line, we introduce a general two-particle state:
\begin{eqnarray}
|\Psi\rangle =\frac{1}{2}\sum_{x_1 \sigma_1,x_2 \sigma_2} \psi_{\sigma_1 \sigma_2}(x_1,x_2)|x_1 \sigma_1, x_2 \sigma_2\rangle,
\end{eqnarray}
which is superposition of Slater determinants $|x_1 \sigma_1, x_2 \sigma_2\rangle=c_{x_2\sigma_2}^{\dagger}c_{x_1 \sigma_1}^{\dagger}|0\rangle$ describing two particles with spin projection $\sigma_1$ and $\sigma_2$ and located at the lattice positions $x_1$ and $x_2$, respectively. Within the mentioned framework, $\psi_{\sigma_1 \sigma_2}(x_1,x_2)$ is the first quantization wavefunction, while $|0\rangle$ represents the empty lattice state. Furthermore, the normalization condition $\langle \Psi| \Psi\rangle=1$ implies:
\begin{eqnarray}
\sum_{x_1 \sigma_1,x_2 \sigma_2} |\psi_{\sigma_1 \sigma_2}(x_1,x_2)|^2=2.
\end{eqnarray}
The requirement that $|\Psi\rangle$ is an eigenstate of the Hamiltonian $\mathcal{H}$ given in Eq. (\ref{eq:Htree}) allows to write the stationary Schr\"{o}dinger equation $\mathcal{H}|\Psi\rangle=E|\Psi\rangle$, being $E$ the energy eigenvalue. Projecting the Schr\"{o}dinger equation on a single Slater determinant, namely $|y_1 s_1, y_2 s_2\rangle$, one obtains:
\begin{eqnarray}
\label{eq:Sc}
&&\langle y_1 s_1, y_2 s_2|H_{kin}|\Psi\rangle+\langle y_1 s_1, y_2 s_2|H_{I}|\Psi\rangle=\nonumber\\
&=&E\psi_{s_1 s_2}(y_1,y_2),
\end{eqnarray}
where $H_{kin}$ represents the kinetic part of the Hamiltonian given in Eq. (\ref{eq:Htree}), while $H_I$ represents the pairing interaction. Direct computation shows that:
\begin{eqnarray}
\label{eq:kinH}
&&\langle y_1 s_1, y_2 s_2|H_{kin}|\Psi\rangle=(\epsilon_{y_1}+\epsilon_{y_2})\psi_{s_1 s_2}(y_1,y_2)+\nonumber\\
&-&K \sum_{y}[\mathcal{A}_{y_{2}y}\psi_{s_1 s_2}(y_1,y)+\mathcal{A}_{y_{1}y}\psi_{s_1 s_2}(y,y_2)],
\end{eqnarray}
being the latter expression clearly affected by the graph connectivity described by adjacency matrix elements $\mathcal{A}_{ij}$. Considering tridiagonal form of the adjacency matrix, i.e. $\mathcal{A}_{ij}=\delta_{i-1,j}+\delta_{i+1,j}$, in Eq. (\ref{eq:kinH}), the usual structure of a two-particle hopping on a linear chain is easily recovered.\\
When the interaction part is considered, different results are obtained depending on the considered pairing model. Different results share a common feature: Interaction only couples particles with opposite spin projection ($s_1 \neq s_2$), while it is ineffective otherwise. Thus, hereafter we focus on the relevant case $s_{1} \neq s_{2}$. In particular, when the attractive Hubbard interaction is considered, one obtains:
\begin{eqnarray}
\label{eq:intHub}
\langle y_1 s_1, y_2 s_2|H_{I}|\Psi\rangle=-g \delta_{y_{1}y_{2}} \psi_{s_{1}s_{2}}(y_{1},y_{2}),
\end{eqnarray}
while a rather different result is derived for the BCS-like pairing, i.e.
\begin{eqnarray}
\label{eq:intBCS}
\langle y_1 s_1, y_2 s_2|H_{I}|\Psi\rangle=-g \delta_{y_{1}y_{2}} \sum_{y}\psi_{s_{1}s_{2}}(y,y).
\end{eqnarray}
Equation (\ref{eq:Sc}) complemented by Eq. (\ref{eq:kinH}), Eq. (\ref{eq:intHub}) or Eq. (\ref{eq:intBCS}) univocally defines the stationary Schr\"{o}dinger equation for the two-body wavefunction $\psi_{s_1 s_2}(y_1,y_2)$ with $s_1 \neq s_2$. Moreover, since the Hamiltonian given in Eq. (\ref{eq:Htree}) preserves the total number of particles $\hat{N}$ and the number $\hat{N}_{\sigma}$ of particles with spin $\sigma \in \{\uparrow,\downarrow\}$ (i.e. $[\mathcal{H},\hat{N}]=[\mathcal{H},\hat{N}_\sigma]=0$), the two-particle problem described by $\psi_{s_1 s_2}(y_1,y_2)$ with $s_1 \neq s_2$ is not coupled with the equal spin problem ($s_1=s_2$) because different spin sectors are independent.\\
The wavefunction symmetry also dictates that fermions in a triplet state with total spin $S=1$ and $S^{z}=0$ are not paired by the considered interaction, while fermions in a singlet state ($S=0$ and $S^{z}=0$) are prone to the pairing interaction. Thus, in order to study interaction effects, the stationary Schr\"{o}dinger equation in Eq. (\ref{eq:Sc}) have to be solved under the requirement that $\psi_{s_1 s_2}(y_1,y_2)=\psi_{s_1 s_2}(y_2,y_1)$, which implies that fermions are in a singlet state.

\subsection{Two-body problem with translational invariance}
Before treating the problem of two interacting fermions in structures lacking of translational symmetry, it is rather instructive to present the translational invariant case. In particular, we study two fermions in a singlet state interacting on a one dimensional lattice. This problem has been already discussed in literature for Hubbard interaction\cite{weisz}, which is here reported as a preparatory problem to the BCS-like case.\\
In the absence of onsite potentials ($\epsilon_i=0$), stationary Schr\"{o}dinger equation of attractive Hubbard model is then written as:
\begin{eqnarray}
\label{eq:Htwo}
(\Delta_1+\Delta_2)\phi(x_1,x_2)-g \delta_{x_1 x_2}\phi(x_1, x_2)=E \phi(x_1,x_2),
\end{eqnarray}
where $\phi(x_1,x_2)$ represents the orbital part of the wavefunction $\psi_{s_1 s_2}(x_1,x_2)$, while the action of the operators $\Delta_1$ and $\Delta_2$ is defined according to the following relations:
\begin{eqnarray}
\Delta_1\phi(x_1,x_2)&=& \! \!   -K \Big [ \phi(x_1-1,x_2)+\phi(x_1+1,x_2)\Big ]\nonumber\\
\Delta_2\phi(x_1,x_2)&=& \! \!   -K \Big [ \phi(x_1,x_2-1)+\phi(x_1,x_2+1)\Big ].\nonumber
\end{eqnarray}
Furthermore, the singlet state requires that $\phi(x_1,x_2)$ is symmetric under particles coordinates exchange. In view of the translational invariance, the stationary Schr\"{o}dinger equation can be solved by means of the ansatz $\phi(x_1,x_2)=e^{ip(x_1+x_2)}f(x_1-x_2)$, being $p$ related to the center of mass momentum. Using the ansatz in Eq. (\ref{eq:Htwo}), we obtain:
\begin{eqnarray}
\label{eq:HtwoReduced}
-2K \cos(p) \Big [ f(z- \! 1)+ \! \! f(z+ \! 1) \Big ] \! - \! g \delta_{z0}f(z)= \! E f(z),
\end{eqnarray}
with $z=x_1-x_2$ and $f(z)=f(-z)$. Using the educated guess $f(z)=f_0 \rho^{|z|}$ with $0<\rho<1$ in Eq. (\ref{eq:HtwoReduced}), we obtain:
\begin{eqnarray}
E_p &=& -\sqrt{g^2+16K^2 \cos^2(p)}\nonumber\\
\rho &=& \sqrt{\frac{|E_p|-g}{|E_p|+g}},
\end{eqnarray}
where $p$ belongs to the interval $(-\pi/2,\pi/2)$ in order to have $\rho>0$. Thus, the paired states form the energy band $-\sqrt{g^2+16 K^2}\leq E_p \leq -g$, while a second band with energy belonging to the interval $(-4K,4K)$ is formed by unpaired states whose wavefunctions are plane waves perturbed by the scattering effect of the pairing potential. The energy bands defined by paired and extended states present a finite overlap except for $g>4K$.\\
An analogous problem can be solved in the case of the BCS-like interaction. In the latter case, the Schr\"{o}dinger equation takes the peculiar form:
\begin{eqnarray}
\label{eq:HtwoBCS}
(\Delta_1+\Delta_2)\phi(x_1,x_2)-g \delta_{x_1 x_2} \! \sum_{y} \! \phi(y, y) \! = \! E \phi(x_1,x_2),
\end{eqnarray}
which is structurally different from Eq. (\ref{eq:Htwo}) and represents one of the main achievements of the present work. Indeed, considering the continuous limit, Eq. (\ref{eq:HtwoBCS}) becomes an integro-differential problem, whose solution could involve mathematical subtilities [\onlinecite{IntegroDifSchr}]. One possible issue is for instance related to the convergence of $\sum_y \phi(y,y)$, which is not ensured when infinite systems are considered. The latter issue is clearly irrelevant for finite size systems and for condensed matter problems where periodic boundary conditions can be imposed.\\
Thus, assuming periodic boundary conditions and using the trial wavefunction $\phi(x_1,x_2)=e^{ip(x_1+x_2)}f(x_1-x_2)$ in Eq. (\ref{eq:HtwoBCS}), we get:
\begin{eqnarray}
\label{eq:HtwoBCSbis}
&&-2K \cos(p) \big [ f(z-1)+f(z+1)\big ]+\nonumber\\
&&-g \delta_{z 0}e^{-ip(x_1+x_2)}\sum_y e^{2ipy}f(0)=E f(z).
\end{eqnarray}
Periodic boundary conditions for a $N$-site system imply that $pN=2\pi m$, with $m \in \mathbb{Z}$, so that $\sum_{y=1}^{N} e^{2ipy}=N \delta_{p0}$. In view of the above arguments, we conclude that BCS-like interaction is effective only for $p=0$, which is an expected feature of the $s$-wave superconductivity. Thus, setting $p=0$ in Eq. (\ref{eq:HtwoBCSbis}), one obtains:
\begin{eqnarray}
\label{eq:HtwoBCSter}
-2K \big [ f(z-1)+f(z+1)\big ]-G \delta_{z0}f(0)=E f(z),
\end{eqnarray}
with $G=gN$ the effective pairing strength. Equation (\ref{eq:HtwoBCSter}) can be solved by means of the same ansatz used in Eq. (\ref{eq:Htwo}), i.e. by setting $f(z)=f_0 \rho^{|z|}$ with $0<\rho<1$. After straightforward algebra, we obtain:
\begin{eqnarray}
\label{eq:expTI}
E_0 &=& -\sqrt{G^2+16K^2}\nonumber\\
\rho &=& \sqrt{\frac{|E_0|-G}{|E_0|+G}}.
\end{eqnarray}
Thus, the BCS-like interaction forms a unique bound state whose energy $E_0$ never overlaps with the energy band formed by unpaired particles. Interestingly, the depairing energy, i.e. the minimum amount of energy required to separate the paired fermions, is given by $\sqrt{G^2+16K^2}-4K$ as long as the thermodynamic limit ($N\gg 1$) is considered, while finite-size effects are expected otherwise.

\begin{figure*}[t!]
\includegraphics[scale=1.30]{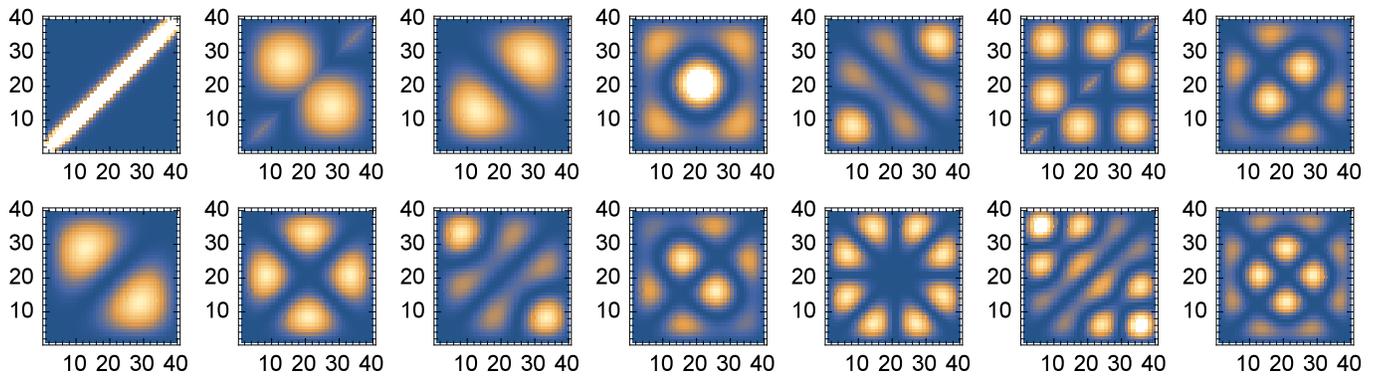}
\caption{Symmetric (upper row) and antisymmetric (lower row) wavefunctions $\phi(i,j)$ of the two-body problem with BCS-like interaction. A linear chain with $N=40$ lattice sites has been considered. Each panel represents $|\phi(i,j)|^2$ as a function of the particles coordinates $i$ and $j$. Darker (clearer) regions represent lower (higher) probability values. Energy eigenvalues of the wavefunctions belonging to the upper (lower) row, ordered from the left to the right, are: $-4.44353$, $-3.97345$, $-3.97069$, $-3.94152$, $-3.92395$, $-3.922$, $-3.90092$ ($-3.97069$, $-3.94152$, $-3.92395$, $-3.90092$, $-3.88335$, $-3.85418$, $-3.84914$). The interaction parameter has been fixed to $g=0.05$. All energy values are expressed in unit of the hopping constant $K$. The wavefunction of the unique paired state is evident looking at the leftmost panel belonging to the upper row.}
\label{figure1}
\end{figure*}

\subsection{Two-body problem with BCS-like pairing: Results for a finite linear chain.}
The simplest system which breaks the translational invariance is a linear chain with finite number $N$ of lattice sites. Equation (\ref{eq:HtwoBCS}), complemented by appropriate boundary conditions, can be numerically solved also implementing the required symmetry of the two-particle wavefunction under coordinates exchange. The numerical procedure implements the simultaneous diagonalization of the two-particle Hamiltonian and of the coodinates exchange operator $\mathcal{P}_x$, acting on the wavefunction as $\mathcal{P}_x \phi(x_1,x_2)=\phi(x_2,x_1)$. In this way, eigenfunctions of the Hamiltonian with specified parity under coordinates exchange are obtained. Since the Hubbard interaction has been the object of intense investigation \cite{hubbardbook}, hereafter we focus on the BCS-like pairing. In particular, we study the problem of two particles subject to the BCS-like pairing and constrained to move on a linear chain with $N=40$ lattice sites. The results of this analysis are presented in Fig. \ref{figure1}, where symmetric (upper row) and antisymmetric (lower row) wavefunctions are obtained by setting a rather strong value of the interaction parameter, i.e. $g=0.05$. The antisymmetric states (lower row) are not paired by the interaction because their wavefunctions are vanishing when the coordinates of the two particles coincide. On the other hand, symmetric states (upper row of Fig. \ref{figure1}) are sensitive to the BCS pairing. In particular, the wavefunction of the unique paired state is evident looking at the leftmost panel belonging to the upper row. The wavefunction of the paired state decreases exponentially as the distance $|x_1-x_2|$ between the two particles increases. Thus, the lack of translational invariance does not perturb too much the translational-invariant picture given in Eq. (\ref{eq:expTI}). Furthermore, the paired state eigenvalue is well approximated by $E_0$ in Eq. (\ref{eq:expTI}) with $G=2$ (notice that $G=gN$ with $g=0.05$ and $N=40$). These observations support the constitutive assumption on which the pairing Hamiltonian for graph-like structures (see Eq. (\ref{eq:Htree})) is based.

\begin{figure}[h!]
\includegraphics[scale=1.10]{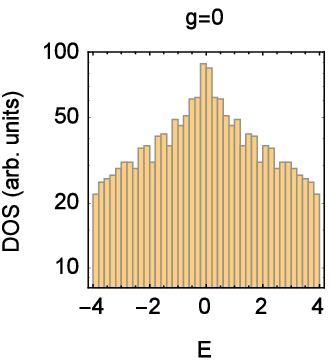}
\includegraphics[scale=1.10]{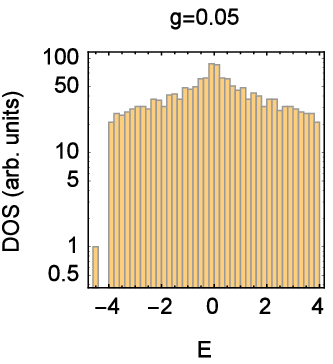}
\caption{Density of states  (in logarithmic scale) of two particles interacting via BCS-like pairing and constrained to move on a one dimensional lattice with $40$ sites. In the absence of interaction (left panel), the density of states is identical to that of a single particle constrained to move on a bidimensional region ($40 \times 40$ square lattice). When finite interaction is considered (right panel), a new subband emerges. The latter, differently from the Hubbard case, contains a unique bound state. For attractive interaction, the bound state energy is located below the bottom of the band formed by unpaired states. In both panels, energies are expressed in units of the hopping $K$.}
\label{figure2ab}
\end{figure}
\begin{figure}[t]
\includegraphics[scale=1.1]{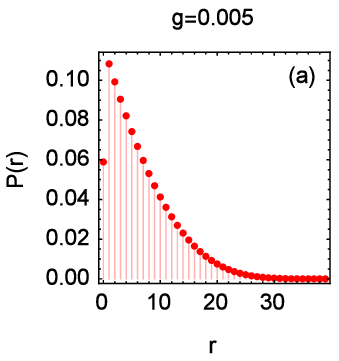}
\includegraphics[scale=1.1]{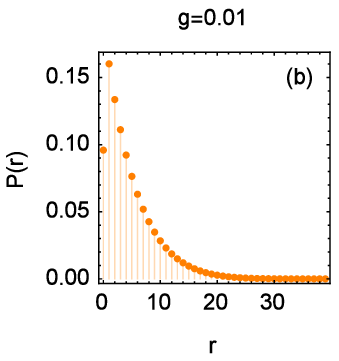}\\
\includegraphics[scale=1.1]{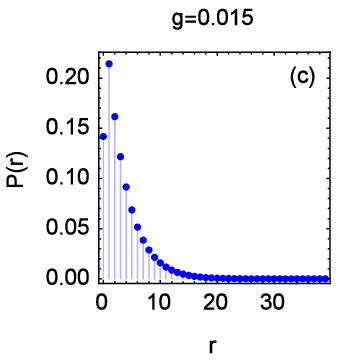}
\includegraphics[scale=1.1]{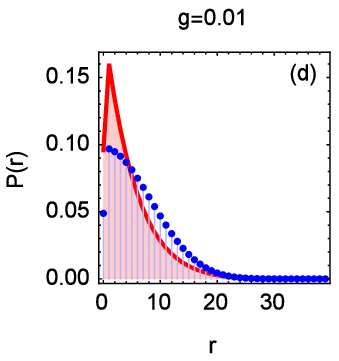}\\
\includegraphics[scale=1.1]{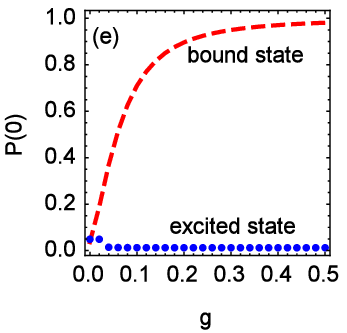}
\caption{Probability distribution $P(r)$ of the particle distance $r$. In panels (a), (b) and (c), $P(r)$ is obtained by considering two fermions forming a bound state (paired electrons); distinct values of the interaction strength, namely $g=0.005$, $0.01$, $0.015$, are considered (see figures labels). In panel (d), a comparison is performed between the probability distribution $P(r)$ pertaining to the bound state (full line) and that of the first excited state (symbols) for $g=0.01$. The probability of finding fermions at the same position, measured by $P(0)$, is an increasing function of the interaction strength $g$ for a two-particle bound state (dashed line of panel (e)), while it is quite insensitive to the interaction when the first excited state is considered (symbols in panel (e)).}
\label{figure3ad}
\end{figure}

The energy spectrum of the two-particle problem can be characterized by studying the density of states (DOS) of particles constrained to move on a one dimensional lattice with $40$ sites. The DOS of the two-particle problem with BCS-like pairing is presented in Fig. \ref{figure2ab}. In the absence of interaction (left panel), the DOS is identical to that of a single particle constrained to move on a bidimensional region ($40 \times 40$ square lattice). When finite interaction is considered (right panel), a new subband emerges. The latter, differently from the Hubbard case, contains a unique bound state. For attractive interaction, the bound state energy is located below the bottom of the band formed by unpaired states. A paired state also exists for repulsive BCS-like interaction, being the energy of this state located above the band formed by the unpaired states. Two-particle bound states stabilized by the interplay between repulsive interaction and spatial discreteness are not unusual and, for instance, they have been extensively described within the framework of the Hubbard model\cite{weisz,souza,deuchert}. Moreover, existence and stability of repulsively bound pairs has been experimentally demonstrated using ultracold bosonic atoms in optical lattices [\onlinecite{zoller2006}]. While the latter findings demonstrate the strong correspondence between the optical lattice physics of ultracold bosonic atoms and the Bose-Hubbard model, the stability of repulsively bound pairs is not expected in condensed matter systems where the simultaneous presence of several degrees of freedom activates efficient decay channels.\\
The two-particle bound state induced by the BCS-like pairing can be further characterized by studying the probability distribution
\begin{eqnarray}
P(r)=\frac{\sum_{ij}|\phi(i,j)|^2\delta_{r,|i-j|}}{\sum_{ij}|\phi(i,j)|^2}
\end{eqnarray}
of the discrete stochastic variable $r$, which measures the distance between two particles belonging to the quantum state $\phi(i,j)$. The probability distribution $P(r)$ is normalized to one according to the relation $\sum_{r \in \{0,..., N-1\}}P(r)=1$. Once $P(r)$ is known, it can be used to compute relevant statistical momenta, i.e. $\langle r^n \rangle=\sum_r r^n P(r)$.\\
Alternatively, the behavior of the $P(r)$ curves can be analyzed to get important information about the spatial organization of fermions belonging to an assigned two-particle quantum state. The mentioned analysis is performed in Fig. \ref{figure3ad}. In particular, in Fig. \ref{figure3ad} (a)-(c), we report $P(r)$ curves obtained by considering electrons forming a bound state (paired electrons) and distinct values of the interaction strength $g$. The $P(r)$ curves present an exponential decay with the fermions distance $r$, which is a behavior reminiscent of the exponential suppression of the paired state wavefunction with the fermions distance. The exponential decay of $P(r)$ is controlled by the interaction strength $g$. As expected, the probability of finding fermions at the same position, measured by $P(0)$, is an increasing function of the interaction strength. Interestingly, within the considered interaction range, $P(r)$ curves are maximized for $r=1$, i.e. for fermions located at adjacent lattice sites.\\
The intuition suggests that the probability distribution of a paired state is substantially different from that of an excited state because the latter is formed by unpaired fermions delocalized along the system. This expectation is corroborated by the analysis performed in Fig. \ref{figure3ad} (d) where the aforementioned comparison is performed considering the paired state and the first excited state. In particular, Fig. \ref{figure3ad} (d) clearly shows that the probability $P(r\leq 3)$ of finding fermions at a distance $r\leq 3$ is equal to $ \sim 0.50$ for a paired state, while the lower value of $\sim 0.33$ is obtained for the excited state.

Based on the above findings, a new question arises about the minimum interaction value required to create a point-like composite particle. The realization of the latter would require that the composite particle formed by the paired fermions lacks spatial extension, being the mentioned condition realized when $P(0) \approx 1$. A close inspection of Eq. (\ref{eq:HtwoBCS}) suggests that a composite point-like particle is trivially realized when the $g\rightarrow \infty$ limit is considered. Under this extreme condition the composite wavefunction is simply given by $\phi_{\infty}(x_1,x_2) \propto \delta_{x_1, x_2}$ with associated energy eigenvalue $E^{\infty}_0 = -G$. The same conclusion can be reached by taking the $G\rightarrow \infty$ limit in Eq. (\ref{eq:expTI}).\\
Apart from the extreme case mentioned above, compact composite particles can be obtained considering finite interaction values. The latter statement is corroborated by Fig. \ref{figure3ad} (e) where the probability of finding fermions at the same position, measured by $P(0)$, is studied as a function of the interaction strength $g$. For a two-particle bound state (dashed line of panel (e)) $P(0)$ is an increasing function of $g$, while it is quite insensitive to the interaction when the first excited state is considered (symbols in panel (e)). In particular, rather compact composite particles are formed when the interaction takes values $g\gtrsim 0.3$ for which $P(0)\gtrsim 0.95$. The interaction values required to obtain a compact composite particle are rather strong in comparison to the hopping energy scale. For instance, the interaction value $g=0.3$ implies an effective interaction parameter $G=12$, in view of the relation $G=gN$ with $N=40$.

\begin{figure}[h]
\includegraphics[scale=0.3]{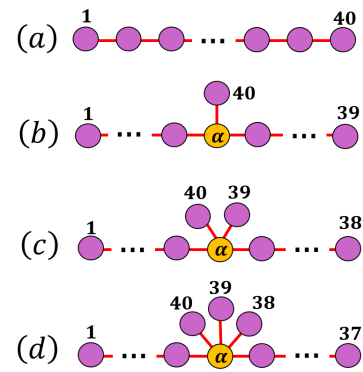}
\caption{(a) Linear chain with $N=40$ lattice sites. (b) Linear chain perturbed by a single lateral site labelled by the index $\alpha$. Linear chains perturbed by two (panel (c)) and three (panel (d)) lateral sites are also shown. As for the panels (b), (c) and (d), different positions of the $\alpha$ site along the chain can be considered. In all the panels, the total number of lattice sites is fixed to $N=40$.}
\label{figure4}
\end{figure}

\subsection{Two-body problem with BCS-like pairing: Results for linear chains perturbed by lateral sites.}
Despite the formalism described so far is suitable to describe arbitrary quantum graphs, hereafter we focus on the simple and relevant case of a linear chain perturbed by the presence of side-sites (see Fig. \ref{figure4} (b)-(d)). In these structures, the the most connected nodes act as localization centers for the particles \cite{deluca}. The localization phenomena induced by the network topology are similar to those induced by the presence of a single-particle potential and can modulate the influence of the pairing mechanism. Thus, the question arises whether the network topology could affect the stability of the two-particle bound state. In order to address this interesting question, we study the \textit{depairing energy} and the bound state wavefunction by considering distinct network topologies. The depairing energy represents the minimum amount of energy to be provided to the two-particle bound state in order to obtain unpaired particles. The latter can be determined directly by measuring the difference between the energy eigenvalues of the first excited state and the ground state. Thus, either the depairing energy and the ground state wavefunction  are obtained by means of the diagonalization procedure.\\
\begin{figure}[!h]
\includegraphics[scale=0.9]{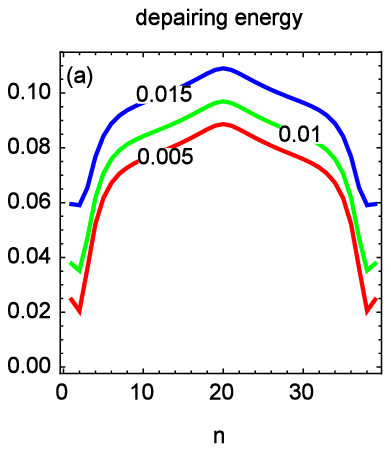}
\includegraphics[scale=0.9]{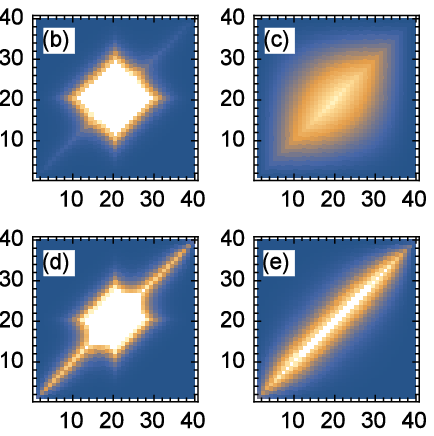}
\caption{Panel (a): Depairing energy of the system depicted in Fig. \ref{figure4}(b) as a function of the position $n \in \{1,...,39\}$ of a single side-site. Different curves are obtained by setting distinct values of the interaction, i.e. $g=0.005$ (lower curve), $g=0.01$ (middle curve) and $g=0.015$ (upper curve). The depairing energy depends on the quantum graph topology and it is maximized when the side-site is located at $n=20$. The modulus squared of the ground state wavefunction of the two-particle problem is shown in panels (b)-(e). In particular, panels (b) and (c) have been obtained by fixing $g=0.005$ and $n=20$ or $n=1$, respectively. Panels (d) and (e) have been obtained by fixing $g=0.015$ and $n=20$ or $n=1$, respectively. Interestingly, the most connected site acts as a localization center for the particles.}
\label{figure5ae}
\end{figure}
In Fig. \ref{figure5ae} (a) we report the depairing energy of the system depicted in Fig. \ref{figure4}(b) as a function of the position $n \in \{1,...,39\}$ of a single side-site. Different curves in panel (a) of Fig. \ref{figure5ae} are obtained by setting distinct values of the interaction, i.e. $g=0.005$ (lower curve), $g=0.01$ (middle curve) and $g=0.015$ (upper curve). For all the curves, the depairing energy depends on the quantum graph topology and it is maximized when the side-site is located at $n=20$. In general, the presence of a side-site at $n \in [10,30]$ appears to be rather beneficial for the two-particle bound state stability. On the other hand, the impact on the bound state stability is less effective when the side-site is located in close vicinity of the system's edges (i.e. $n \in [1,9]$ or $n \in [31,39]$). Indeed, under the latter condition, the confinement effects induced by the side-site are altered by the presence of one edge of the system. The reflection symmetry of the depairing energy curves around the mirror line $n=20$ is reminiscent of the invariance of the system properties under appropriate relabelling of the network nodes. A remarkable example of the above observation is the equivalence of systems with a side-site located at $n=1$ or $n=39$, respectively. Indeed, after nodes relabelling, the aforementioned systems are mapped into the linear chain shown in Fig. \ref{figure4} (a). This observation provides us the opportunity to perform a quantitative comparison between the depairing energies of a linear chain (obtained by setting $n=1$ or $n=39$) and the perturbed linear chain with a side-site located at $n=20$. In this respect, we introduce the auxiliary variable $\eta(g)=\Delta_{20}(g)/\Delta_{1}(g)$ measuring the ratio between the depairing energy $\Delta_{20}$ of a linear chain with a side-site located at $n=20$ and the depairing energy $\Delta_{1}$ of a linear chain. An inspection of Fig. \ref{figure5ae} (a) shows that the ratio $\eta(g)$ is greater than one within the interaction range considered, while it is a decreasing function of the interaction strength $g$. In particular, we find $\eta(0.005) \approx 3.60$, $\eta(0.01) \approx 2.57$ and $\eta(0.015) \approx 1.83$, being the latter sequence a clear indication of the non-trivial interplay between confinement effects (controlled by the network topology) and two-body interaction. This statement can be verified by looking at the morphology of the bound state wavefunction, which, on its turn, is affected by the network topology.\\
The modulus squared of the ground state wavefunction of the two-particle problem is shown in panels (b)-(e) of Figure \ref{figure5ae}. Panels (b) and (c) have been obtained by fixing $g=0.005$. In particular, panel (b) represents the modulus squared of the ground state wavefunction of two particles confined to a linear chain perturbed by a side-site located at $n=20$, while the case of a linear chain is shown in panel (c). When the case of a linear chain is considered (panel (c))), the two particle probability density shows the tendency to localize along the line $x_1=x_2$, being $x_i$ the coordinate of the i-th particle. The latter behavior is evidently promoted by the two-body attractive interaction. On the other hand, the presence of a side-site at $n=20$ (see panel (b)) acts as a single-particle potential which forces each particle to stay in close vicinity of $n=20$.  As a consequence, in the latter case, the variance $\sigma^{2}_X$ of the center of mass coordinate $X=(x_1+x_2)/2$ of the two-particle bound state is reduced compared to the case shown in panel (c). The $\sigma^{2}_X$ reduction, controlled by the network topology, is accompanied by a simultaneous enhancement of the depairing energy. Thus, a reduced variance of the center of mass coordinate seems to be a favorable condition for the bound state stability. This interpretation is also consistent with the analysis of panels (d) and (e), where the interaction strength has been fixed to $g=0.015$. Panel (e) shows the case of a linear chain where the two-body interaction, which is stronger compared to panel (c), produces a cigar-shaped particle density. When the side-site case is considered (see panel (d)), one observes that the effective single particle potential produced by the network topology tends to attract the particles in close vicinity of the side site even though it is not able to depopulate the particle density distribution tails located along the $x_1=x_2$ line. In general, the center of mass confinement, signaled by a reduction of $\sigma^{2}_X$, indirectly affects the dynamics of the relative coordinate $x=x_1-x_2$, since these degrees of freedom are not independent in the presence of confinement. Confinement effects on the relative coordinate $x$ can be quantified by studying the coherence length $\xi_C$. The outcome of this analysis, reported in appendix \ref{app:A}, shows that the coherence length is reduced in the presence of a side site. Thus, the depairing energy enhancement is accompained by coherence length reduction, which is expected within a BCS picture.\\
The aforementioned arguments are also applicable to the networks shown in Fig. \ref{figure4} (c) and (d), where linear chains with two or three side-sites, respectively, are depicted. In these systems the strength of the effective single particle potential produced by the network topology is a growing function of the number of side-sites. Since the affectiveness of the topological confinement of the particle density grows with the number of side-sites, the depairing energy of these structures presents a strong enhancement compared to the case of a linear chain (Fig. \ref{figure4} (a)).
\begin{figure}[h]
\includegraphics[scale=0.91]{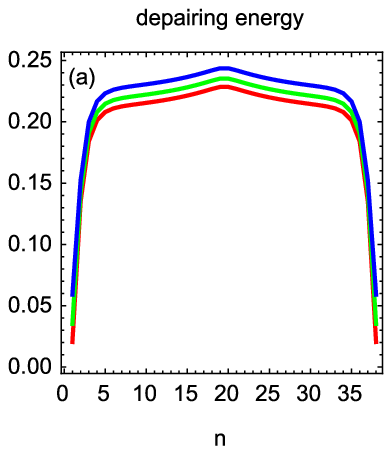}
\includegraphics[scale=0.89]{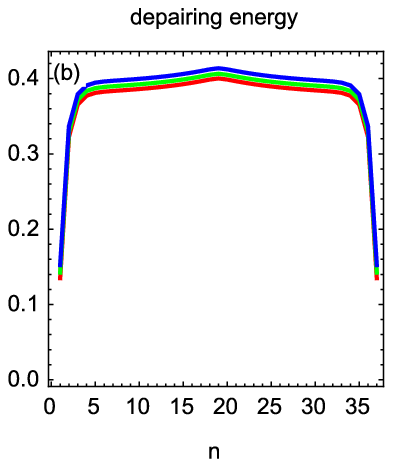}
\caption{Panel (a): Depairing energy as a function of the position $n\in \{1,..., 38\}$ of the side-sites for the network shown in Fig. \ref{figure4} (c); Panel (b): Depairing energy as a function of the position $n \in \{1,..., 37\}$ of the side-sites for the network shown in Fig. \ref{figure4} (d). For both the panels, different curves are obtained by setting distinct values of the interaction, i.e. $g=0.005$ (lower curve), $g=0.01$ (middle curve) and $g=0.015$ (upper curve).}
\label{figure6ab}
\end{figure}
The latter observation can be verified by looking at Fig. \ref{figure6ab} where the depairing energy of a linear chain with two (Fig. \ref{figure6ab} (a)) or three (Fig. \ref{figure6ab} (b)) side-sites is monitored as the location $n$ of the side-sites is moved along the chain. Similarly to what observed in Fig. \ref{figure5ae} (a), the depairing energy curves in panel (a) or (b) of Fig. \ref{figure6ab} show a maximum when the side-sites are located in the middle of the chain. The depairing curves are almost insensitive to the location of the side-sites provided that they are sufficiently distant from the chain ends. For the interaction range considered, the maximum values of the depairing energies in both the panels are rather insensitive to the specific choice of the interaction $g$, while a sensitive dependence on the number of side-sites is observed. The magnification of the depairing energy induced by the network topology can be quantified by comparison with the linear chain case. For instance, setting $g=0.015$, the depairing energy for the linear chain case is $\approx 0.06$ in units of $K$, while the maximum depairing energy for a linear chain with two side-sites (panel (a)) is $\approx 0.24$. Moreover, a linear chain with three side-sites (panel (b)) exhibits the value $\approx 0.42$. The above estimates show that in networks like those depicted in Fig. \ref{figure4} (c)-(d) the stability of the two-particle bound state, measured through the depairing energy, is enhanced by a factor ranging from $4$ to $7$ in comparison to the linear chain case (Fig. \ref{figure4} (a)).

\section{Few-body problem}
\label{sec:many}

The analysis of the two-body problem led us to the conclusion that the stability of the two-fermion bound state is enhanced in linear chains perturbed by side-sites. The enhancement effect originates from the confining properties of network's sites with higher connectivity, which behave like attractive single-particle potentials. However, these localization centers cannot host more than two fermionic particles. Thus, one might ask whether the stabilizing effect of the network connectivity maintains its effectiveness in the presence of many-body effects.

\subsection{Richardson's solution of the many-body problem in quantum graphs}
\begin{figure*}[t!]
\includegraphics[scale=1]{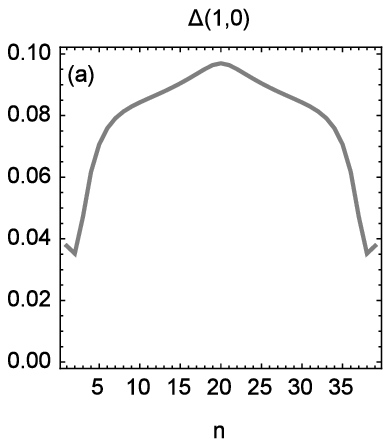}
\includegraphics[scale=1.05]{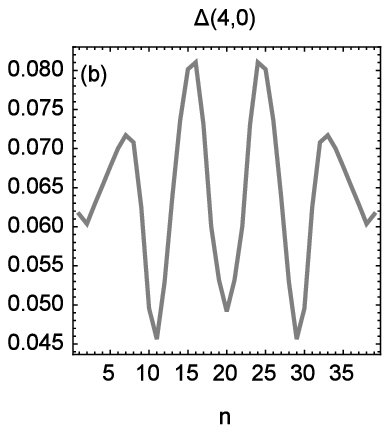}
\includegraphics[scale=1.02]{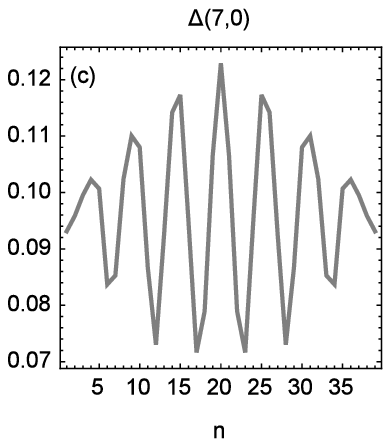}\\
\includegraphics[scale=1.05]{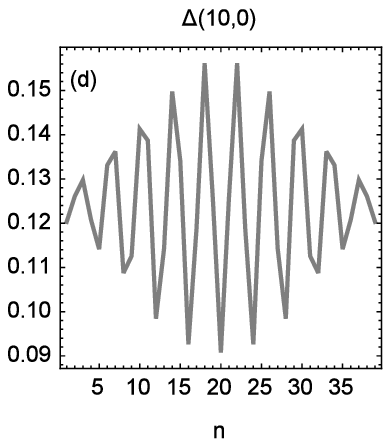}
\includegraphics[scale=1.05]{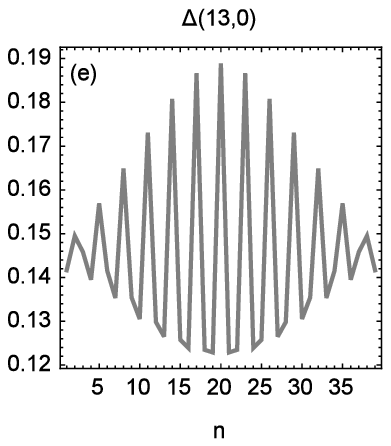}
\includegraphics[scale=1]{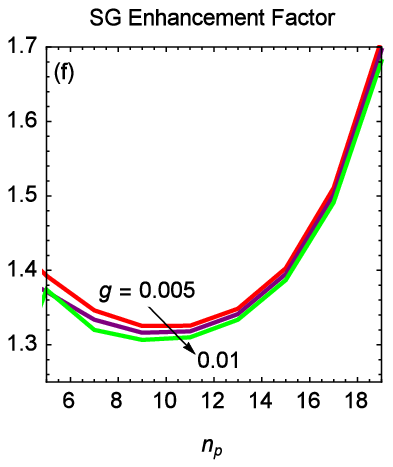}
\caption{Curves of the spectral gap $\Delta(n_p,0)$ as a function of the position $n$ of the side site are reported in panels (a)-(e). Different panels are obtained by considering the network topology shown in Fig. \ref{figure4} (b) and setting $g=0.01$ and $n_p \in \{1,4,7,10,13\}$, as specified in each panel. Panel (f) shows the spectral gap enhancement factor as a function of $n_p$. Different curves are obtained by considering distinct interaction values, i.e. $g=0.005$ (upper curve), $g=0.0075$ (middle curve) and $g=0.01$ (lower curve).}
\label{figure7}
\end{figure*}

In order to address the aforementioned relevant question in a simple way, we consider an even number $\mathcal{M}$ of particles described by the Hamiltonian given in Eq. (\ref{eq:reducedBCS}). The Hamiltonian in Eq. (\ref{eq:reducedBCS}) can be written in terms of the operators $b_{i}$ and $b_i^{\dag}=a^{\dag}_{i\uparrow}a^{\dag}_{i\downarrow}$, being the latter the creation operator of a pair formed by particles occupying the same single-particle energy level $E_i$ and presenting opposite spin orientations. Under the assumption that all sigle-particle energy levels (whose number is equal to the number $N$ of lattice sites) are doubly occupied or empty, the Hamiltonian in Eq. (\ref{eq:reducedBCS}) takes the form:
\begin{eqnarray}
\label{eq:Hrich1}
\mathcal{H}=\sum_{i}2E_i b^{\dag}_{i}b_{i}-g \sum_{\alpha \beta}b^{\dag}_{\alpha}b_{\beta},
\end{eqnarray}
where $i$, $\alpha$ and $\beta$ are indices labelling the single-particle energy levels and running over the entire spectrum (i.e., $i, \alpha, \beta \in \{1,..., N\}\equiv \Omega$ and $E_{i+1}>E_i$, $\forall \ i \in \Omega$). The presence of singly-occupied energy levels provides a limited modification of the Hamiltonian in Eq. (\ref{eq:Hrich1}). Indeed, introducing the set $B$, with cardinality $b$, collecting the indices of the single-occupied energy levels and its complement $\overline{B}$, with cardinality $N-b$, collecting the indices of the doubly-occupied or empty energy levels, one obtains:
\begin{eqnarray}
\label{eq:Hrich2}
\mathcal{H}=\sum_{i \in B}E_i n_i+\sum_{i \in \overline{B}}2E_i b^{\dag}_{i}b_{i}-g \sum_{\alpha \beta \in \overline{B}}b^{\dag}_{\alpha}b_{\beta},
\end{eqnarray}
where $n_i=a^{\dag}_{i\uparrow}a_{i\uparrow}+a^{\dag}_{i\downarrow}a_{i\downarrow}$. Interestingly, the singly-occupied levels are not available to
the pairs scattering due to the Pauli principle (blocking effect) and, for this reason, the labels of such levels, belonging to $B$, are good quantum numbers. Thus, the first term in Eq. (\ref{eq:Hrich2}) just provides an energy shift $\mathcal{E}_B=\sum_{i \in B}E_i$ to the total energy. The dynamics of the remaining $n_p$ pairs is instead governed by the Hamiltonian
\begin{eqnarray}
\label{eq:Hrich3}
\mathcal{H}_{\overline{B}}=\sum_{i \in \overline{B}}2E_i b^{\dag}_{i}b_{i}-g \sum_{\alpha \beta \in \overline{B}}b^{\dag}_{\alpha}b_{\beta},
\end{eqnarray}
whose structure, apart from the limitation over the summation indices, is identical to that of Eq. (\ref{eq:Hrich1}). Despite the innocent form of  Eq.(\ref{eq:Hrich3}), the difficult part of the problem is hidden inside the non-canonical algebra of the $b$ operators, satisfying the hard-core boson relations $(b_j^{\dag})^2=0$ and $[b_{j},b_{j'}^{\dag}]=\delta_{jj'}(1-2 b_{j}^{\dag}b_{j})$. Richardson showed that a generic many-body eigenstate of the Hamiltonian in Eq. (\ref{eq:Hrich2}) can be obtained starting from the vacuum state $|0\rangle$ according to the relation \cite{sierra}:
\begin{eqnarray}
|n_p,B\rangle \propto \prod_{i \in B}a^{\dag}_{i \sigma^{(i)}}\prod_{\nu=1}^{n_p} \Bigl ( \sum_{j \in \overline{B}}\frac{b_j^{\dag}}{2E_{j}-e_{\nu}}\Bigl ) |0\rangle,
\end{eqnarray}
where $\mathcal{M}=2n_p+b$, $\sigma^{(i)} \in \{\uparrow,\downarrow\}$ is the spin projection of a particle in the singly-occupied level labelled by $i$, while $e_{\nu}$ are particular solutions of the set of the $n_p$ coupled algebraic equations:
\begin{eqnarray}
\label{eq:richeq}
1+\sum^{n_p}_{\mu=1 (\neq \nu)}\frac{2g}{e_{\mu}-e_{\nu}}=\sum_{j \in \overline{B}}\frac{g}{2E_j-e_{\nu}},
\end{eqnarray}
with $\nu \in \{1,..., n_p\}$. All possible solutions of Eq. (\ref{eq:richeq}) can be presented in the form $\{e_1^{(\lambda)},...,e_{n_p}^{(\lambda)}\}$, where the maximum value of $\lambda \in \{1,..., \lambda_M \}$ is given by
\begin{eqnarray}
\lambda_M=\frac{(N-b)!}{n_p!(N-b-n_p)!},
\end{eqnarray}
which represents the number of distinct eigenstates with $n_p$ pairs and $b$ singly-occupied levels. Despite the numerical solution of Eq. (\ref{eq:richeq}) is highly non-trivial \cite{richardson2}, solution algorithms can be optimized by observing that generic solutions $e_{\nu}$ appear in close vicinity of $2E_j$ for negligible interaction values and smoothly evolve toward lower values as the interaction increases.\\
The action of the Hamiltonian $\mathcal{H}$ in Eq. (\ref{eq:Hrich2}) on the generic eigenstate  $|n_p,B\rangle$ is specified by $\mathcal{H}|n_p,B\rangle=\mathcal{E}_{b}(n_p)|n_p,B\rangle$ with $\mathcal{E}_{b}(n_p)=\mathcal{E}_B+\sum_{\nu} e_{\nu}$. The eigenstate $|n_p,B\rangle_{o}$ with minimum energy $\mathcal{E}^{o}_{b}(n_p)$, characterized by $n_p$ paired particles and $b$ singly-occupied levels, is obtained by allocating unpaired particles in levels whose energy is positioned as close as possible to the uncorrelated $\mathcal{M}$-particle Fermi level.\\
The minimum energy $\Delta(n_p,b)$ required to break a pair, i.e. the so-called spectroscopic gap, can be computed as\cite{sierra}
\begin{eqnarray}
\Delta(n_p,b)=\mathcal{E}^{o}_{b+2}(n_p-1)-\mathcal{E}^{o}_{b}(n_p).
\end{eqnarray}
Conceptually, $\Delta(n_p,b)$, which is a genuine many-body quantity, plays the same role of the depairing energy introduced for the two-body problem. Indeed, the depairing energy of the two-body problem can be also studied by means of the Richardson approach, being the quantity of interest $\Delta(1,0)$. When the quantity $\Delta(1,0)$ is compared with the depairing energy obtained by solving the two-particle eigenproblem in Eq. (\ref{eq:HtwoBCS}) a full agreement is found. Thus, the spectroscopic gap, which is an implicit function of the network topology, can be used to study the interplay between the non-trivial structure of the lattice (network topology) and many-body effects.\\
In order to follow this program, hereafter we focus our attention on the many-body physics of particles constrained to move on a linear chain perturbed by a single lateral site (see Fig. \ref{figure4} (b)). In particular, we consider a network containing $N=40$ lattice sites, so that $N$ also represents the number of single particle levels. We consider a system with even number $\mathcal{M}=2 n_p$ of particles implying that the many-body ground state $|n_p, \emptyset \rangle_o$ lacks of singly occupied levels (i.e., $b=0$). The first excited state $|n_p-1, B=\{i,j\} \rangle$ is obtained starting from the ground state and redistributing the particles belonging to the highest doubly occupied level into the singly occupied level $E_i$ and $E_j$, being the latter as close as possible to the non-interacting Fermi level. The energy difference between the ground state and the first excited state is measured by the spectral gap $\Delta(n_p,0)$. The latter quantity depends on the number of paired state $n_p$ and also contains non-trivial information about the network topology.\\
A systematic study of the spectral gap $\Delta(n_p,0)$ as a function of the position $n$ of the side site is presented in Fig. \ref{figure7} (a)-(e). The different curves have been obtained by setting the interaction strength to $g=0.01$, while changing the number $n_p \in \{1,4,7,10,13\}$ of the paired states.\\
In Fig. \ref{figure7} (a) we present the space dependence of the spectroscopic gap of a single pair problem, which is the same situation studied in Fig. \ref{figure5ae} (a) (middle curve) by means of the depairing energy. The direct comparison of the two curves shows a complete agreement implying that, as far as the two-particle problem is concerned, the Richardson approach is equivalent to the eigenproblem in Eq. (\ref{eq:HtwoBCS}). Increasing the particles number, as done in Fig. \ref{figure7} (b)-(e), allows a systematic study of the many-body effects. The latter introduce a modulation in the curves of the spectroscopic gap $\Delta(n_p,0)$ as a function of the position $n$ of the side site. In particular, a multi-peak structure with $n_p$ peaks affects the $\Delta(n_p,0)$ \textit{vs} $n$ curves. Since the number of peaks is fixed to $n_p$ and in view of the required reflection symmetry around the mirror line $n=20$, one observes that the $\Delta(n_p,0)$ \textit{vs} $n$ curves present a peak in $n=20$ if $n_p$ is odd, while a local minimum is observed when $n_p$ is even. Interestingly, the multi-peak structure of the $\Delta(n_p,0)$ \textit{vs} $n$ curves seems to be reminiscent of the spatial distribution of the pairs density. Despite the mentioned deformation, the $\Delta(n_p,0)$ \textit{vs} $n$ curves present an envelop with a maximum located at $n=20$. Thus, the spectroscopic gap is enhanced in linear chain perturbed by a well-positioned side site compared to the case of a linear chain (Fig. \ref{figure4} (a)), even in the presence of many-body effects.\\
The spectroscopic gap enhancement factor can be quantified by studying the ratio between the maximum of the spectral gap curves and the spectral gap of a linear chain. The analysis is performed in Fig. \ref{figure7} (f) where the mentioned ratio is reported as a function of $n_p$, being the latter an odd number. The distinct curves in Fig. \ref{figure7} (f), obtained by setting different interaction values, appear to be rather insensitive to the specific interaction strength, at least within the considered parameters range. In particular, we observe an enhancement factor $\sim 1.3-1.4$ for $n_p \leq 15$, while it shows the tendency to increase with the number of pairs for $n_p>15.$ This behavior originates from the variability of the single-particle levels spacing, which plays a relevant role in defining the many-body physics of the problem \cite{sierra}.\\

\subsection{BCS parametrization of the many-body correlations}
\begin{figure*}[t!]
\includegraphics[scale=1.2]{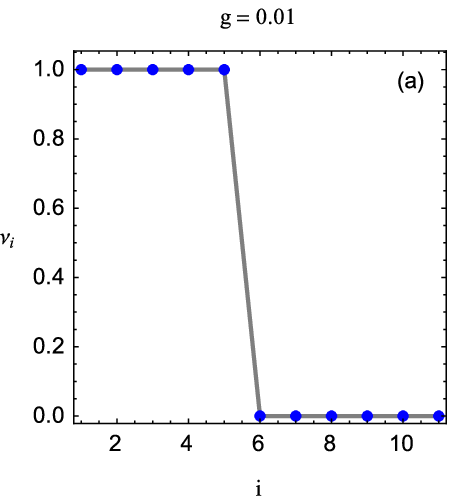}
\includegraphics[scale=1.2]{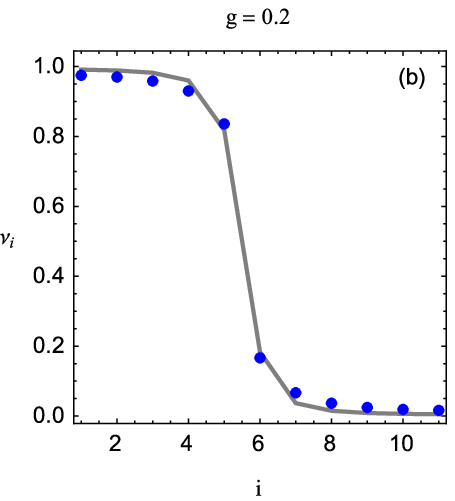}
\includegraphics[scale=1.2]{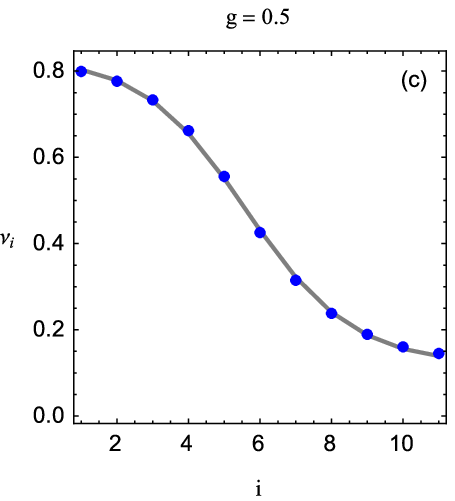}\\
\includegraphics[scale=1.2]{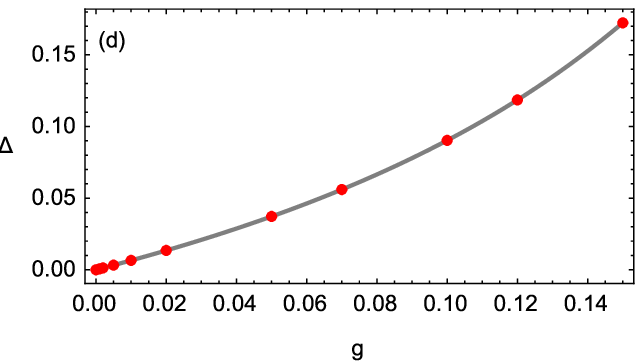}
\includegraphics[scale=1.2]{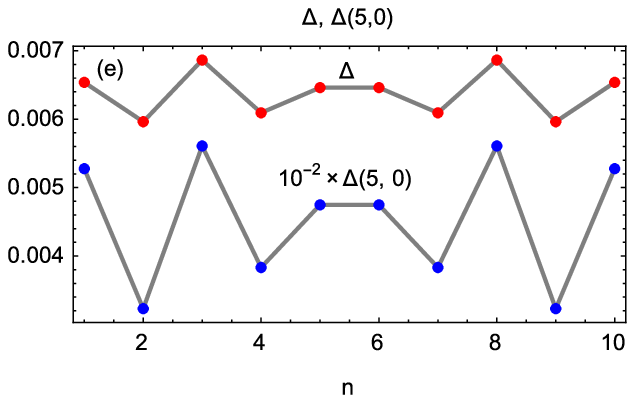}
\caption{Panels (a)-(c): Occupation numbers $\nu_i$ in the many-body ground state of a linear chain network with $N=11$ and $n_p=5$. Symbols are obtained by exact diagonalization of the problem, while full lines represent the fitting curves obtained by using Eq.(\ref{eq:bcsDO}). Panels have been obtained by setting distinct interaction values. Panel (d): Pairing parameter $\Delta$ deduced by the fitting procedure as a function of $g$ (symbols). A linear chain network with $N=11$ and $n_p=5$ has been considered. The full line is the interpolation curve $\Delta(g) = a_1 g+ a_2 g^2+a_3 g^3$ with $a_1 \approx 0.634778$, $a_2 \approx 1.8402$, $a_3 \approx 6.05468$. Panel (e): $\Delta$ and $\Delta(5,0)$ as a function the side site position $n$, assuming a linear chain perturbed by a single lateral site. The system parameters have been fixed as $N=11$, $n_p=5$ and $g=0.01$.}
\label{figure8}
\end{figure*}

An alternative way to quantify the superconducting correlations in the many-body ground state $|n_p,\emptyset\rangle_{o}$ is studying the occupation numbers of each single-particle level. The occupation numbers are proportional to the quantity $\nu_i=_{o}\langle n_p,\emptyset|b^{\dag}_i b_i|n_p,\emptyset\rangle_{o}$, being the spin degeneracy the proportionality factor. In absence of interaction (i.e., for $g=0$), the numbers $\nu_i$ are described by the Heaviside step function $\theta(E_F-E_i)$, which is the zero-temperature limit of the Fermi function describing a Fermi see filled up to the Fermi energy $E_F$. For finite interaction values the step-like function, characteristic of the non-interacting problem at zero temperature, is replaced by a smooth function whose behavior is similar to a Fermi function at finite temperature. However, despite the qualitative similarity, the Fermi function is not adequate to describe the numbers $\nu_i$. Instead, they are well described by the BCS probability to find the level $i$ doubly occupied, namely\cite{vondelft}
\begin{eqnarray}
\label{eq:bcsDO}
v^2_i=\frac{1}{2}\Bigl [ 1-\frac{E_i-\mu}{\sqrt{(E_i-\mu)^2+\Delta^2}}\Bigl],
\end{eqnarray}
where $\mu$ and $\Delta$ are the grand-canonical chemical potential and the superconducting gap of the BCS theory. Despite we are working within a canonical picture, i.e. at fixed particles number, Eq. (\ref{eq:bcsDO}) can be used to reproduce the $\nu_i$ \textit{vs} $i$ behavior once the parameter $\mu$ and $\Delta$ have been fixed by means of a fitting procedure. The latter is rather accurate and works efficiently even when a small number of particles is considered. Moreover, the constraint $\sum_i v^2_i=n_p$ is always respected with high accuracy. The pairing parameter $\Delta$, extracted by the fitting procedure, depends on the interaction strength $g$ and on the network topology. More importantly, $\Delta$ is conceptually and numerically different from the order parameter of the ordinary BCS theory \cite{vondelft}. Although these differences can be neglected only in thermodynamic limit, the many-body correlations can be parameterized by using Eq. (\ref{eq:bcsDO}).\\
In order to corroborate this statement, in Fig. \ref{figure8} (a)-(c) we present the occupation numbers $\nu_i$ in the many-body ground state of a linear chain network with $N=11$ and $n_p=5$. The $\nu_i$ \textit{vs} $i$ curves show a step-like behavior as long as a weak interaction strength is considered (see panel (a)), while a smeared step-function characterizes the curves obtained by setting higher interaction values (see panels (b) and (c)). The aforementioned behavior is well described by Eq. (\ref{eq:bcsDO}) so that a fitting procedure can identify the parameters $\mu$ and $\Delta$. Once the parameter $\Delta$ has been identified, it can be studied as a function of the interaction strength $g$.\\
This analysis, performed in Fig. \ref{figure8} (d), shows that the pairing parameter $\Delta$ is described by a linear function of $g$ as long as the weak interaction limit is considered, while deviations from this behavior are present for increasing interaction values. The non-BCS behavior of the pairing parameter $\Delta$ as a function of the interaction strength $g$ is strongly affected by finite-size effects and it is reminiscent of the analogous behavior of the two-body coherence length (see appendix \ref{app:A}). Once the linear chain network has been characterized, it is possible to study a linear chain perturbed by a single lateral site with $N=11$ and $n_p=5$. In particular, the $\Delta$ \textit{vs} $n$ behavior, with $n$ the position of the lateral site, can be derived.\\
In Fig. \ref{figure8} (e) we show the results of the mentioned analysis by setting $g=0.01$. In particular, the parameter $\Delta$ appears to be sensitive to the network topology and it is enhanced, compared to the linear chain network case, for appropriate locations of the lateral site. The $\Delta$ \textit{vs} $n$ behavior can be compared with the spectroscopic gap dependence on $n$. The two curves present a similar behavior, even though the spectroscopic gap $\Delta(5,0)$ is two orders of magnitude greater then $\Delta$. This difference, which is expected to disappears in thermodynamic limit, is originated by the sensitive dependence of the spectroscopic gap on the level spacing. The latter is greater then $g=0.01$ for the reduced-size system considered and this motivates the different scales of the mentioned curves. Despite these subtilities, the above analysis demonstrates that many-body correlations can be captured, to some extent, by using a BCS parametrization of the ground state properties.

\section{Discussion and conclusions}
\label{sec:concl}
In conclusion, we have presented a rather comprehensive study of the pairing interaction in quantum graphs showing that the network topology plays a relevant role in deciding the effectiveness of the many-body interactions. To reach these conclusions, we have proposed a real-space version of the BCS pairing interaction which allows to study the interplay between pairing and network topology. We have demonstrated that the resulting many-body Hamiltonian can be treated by means of the Richardson method. When the two-body problem is considered, we find that the wavefunction of the problem is solution of an integro-differential Schr\"{o}dinger equation. The latter can be studied numerically in a variety of conditions and can be used to obtain information about the depairing energy, i.e. the minimum amount of energy required to separate the paired particles forming a bound state. We have shown that the depairing energy can be enhanced in networks with appropriate topology and the enhancement phenomenology survives when the many-body problem is considered.\\
These findings are in logical continuity with those reported in Ref. [\onlinecite{micronetdiscreto}] where, by using a modified DeGennes-Alexander micronetwork theory, it has been demonstrated that the superconducting critical temperature in star-like networks is enhanced compared to the transition temperature of a single disconnected island. This behavior, which is experimentally confirmed \cite{lucci3}, suggests that the enhancement of the superconducting gap detected in these systems originates from amplification phenomena of the effective interaction strength induced by the network topology.\\
In view of the generality of our findings, it is expected that selected aspects of the reported phenomenology can be also tested by using \textit{atomtronics} technologies\cite{cold,cold2}.

\section*{Acknowledgment}
The author is indebted to Matteo Cirillo whose experimental activity has motivated the present work. Inspiring discussions with M. Salerno about interacting quantum systems are gratefully acknowledged. Comments by A. Avella, M. Blasone, R. Citro, F. Corberi, R. De Luca, A. Maiellaro, C. Noce and J. Settino about diverse aspects of the present work are also acknowledged. R. De Luca and R. Citro are also acknowledged for comments about a preliminary version of the manuscript.

\section*{Data Availability Statement}
No data associated in the manuscript.

\appendix
\section{Coherence length in quantum graphs}
\label{app:A}
The coherence length is an important parameter of the superconducting state. It can be easily obtained in the context of the two-body problem once the ground state wavefunction $\phi(i,j)$ is known. In particular, the coherence length $\xi_C$, inspired by the definition due to Cooper, is given by:
\begin{eqnarray}
\xi_C^2=\sum_{ij}\mathcal{D}_{ij}^2|\phi(i,j)|^2,
\end{eqnarray}
where $\mathcal{D}_{ij}$ is the length of the shortest path connecting the nodes $i$ and $j$. As long as a linear chain is considered (see Fig. \ref{figure4}(a)), $\mathcal{D}_{ij}=|i-j|$, while a different definition of $\mathcal{D}_{ij}$ is required for graphs like those presented in Fig. \ref{figure4}(b). When a linear chain perturbed by a single lateral site (Fig. \ref{figure4}(b)) is considered, despite a certain degree of arbitrariness is possible, it is reasonable to assume:
\begin{eqnarray}
\mathcal{D}_{ij}=|z_i-z_j|+|\delta_{iN}-\delta_{jN}|,
\end{eqnarray}
with $z_i$ a function taking value $i$ when $i \neq N$ and $\alpha$ otherwise. Moreover, $N$ and $\alpha$ represent the number of lattice sites and the index of the node attached to the lateral site, denoted by the index $N$.\\
\begin{figure}[t!]
\includegraphics[scale=1.2]{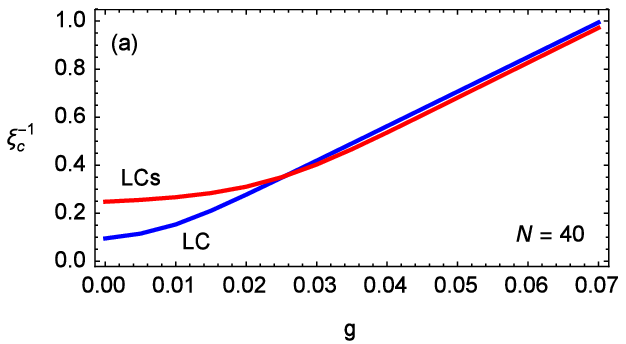}\\
\includegraphics[scale=1.2]{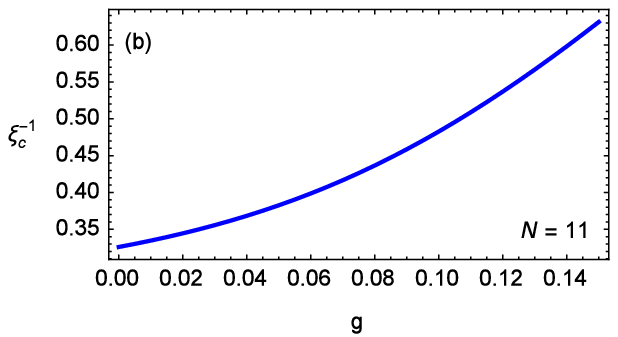}
\caption{Panel (a): $\xi_C^{-1}$ \textit{vs} $g$ curves obtained by considering a linear chain ($LC$) or a linear chain perturbed by a side site located at $n=20$ ($LCs$). The total number of lattice sites is $N=40$ for all the considered cases. Panel (b): $\xi_C^{-1}$ \textit{vs} $g$ curve obtained by considering a linear chain with $N=11$ lattice sites.}
\label{figure9}
\end{figure}
Using the definition of coherence length provided above, we are able to further characterize the physical situation described in Fig. \ref{figure5ae} of the main text. This analysis is presented in Fig. \ref{figure9}(a), where we study $\xi_C^{-1}$ as a function of the interaction strength $g$. In particular, the curve labelled by $LC$ refers to the case of a linear chain, while the curve with label $LCs$ is obtained by considering a linear chain with a side site in position $n=20$. The total number of lattice sites is $N=40$ for all the analyzed cases. The figure clearly shows that the presence of a side site reduces the coherence length compared to the case of a linear chain. The entity of this reduction is controlled by the interaction strength. In the strong interaction limit, no significant difference is observed between the $LC$ and the $LCs$ curves. The mentioned behavior depends on the fact that, increasing the interaction strength $g$, the two-particle bound state presents a reduced extension and therefore the effects of the side site are less pronounced. The main conclusion of the above discussion is that the enhancement of the depairing energy produced by a side site (in comparison with the case of a linear chain) is accompanied by a reduction of the coherence length, being the latter conclusion compatible with the BCS inverse proportionality relation between coherence length and energy gap.\\
In order to understand the results reported in Fig. \ref{figure8}(d), we also study, in Fig. \ref{figure9}(b), the $\xi_C^{-1}$ \textit{vs} $g$ curve obtained by considering a linear chain with $N=11$ lattice sites. The comparison between Fig. \ref{figure8}(d) and Fig. \ref{figure9}(b) suggests that the non-BCS behavior of $\Delta(g)$, as deduced by the fitting procedure described in the main text, is reminiscent of the $\xi_C^{-1}$ \textit{vs} $g$ relation of the two-body problem.\\
Interestingly, the curves in Fig. \ref{figure9} are clearly affected by finite-size effects. Indeed, in the absence of interaction (i.e. $g=0$) and considering the thermodynamic limit (i.e. $N\rightarrow \infty$), it is expected that $\xi_C^{-1}$ be zero. The Curves in Fig. \ref{figure9} do not respect this requirement and one can verify that $LC$ in Fig. \ref{figure9}(a) and the curve in Fig. \ref{figure9}(b) exhibit the finite-size scaling $\xi_C^{-1}(g=0) \approx 2\sqrt{3}N^{-1}$.

\end{document}